\documentclass[preprint,aps,singlecolumn,superscriptaddress,showpacs,floatfix,amssymb,amsmath]{revtex4-1}
\usepackage{epsfig}
\usepackage{graphics}

\usepackage{float} 
\usepackage{array}
\usepackage{multirow}

\usepackage{yfonts}

\usepackage{graphicx}
\usepackage{amssymb}
\usepackage{mathrsfs}
\usepackage{bbm}
\usepackage{epsfig}
\usepackage{makecell}

\usepackage[usenames,dvipsnames]{color}

\usepackage{bbm}
\usepackage{color}

\newcommand{\la}[1]{\mbox{$
\lefteqn{ \mbox{\,\, \tiny #1}}$} \label{#1}}

\newcommand{\be}{\begin{eqnarray}}
\newcommand{\ee}{\end{eqnarray}}

\begin{document}

\title{
Topologically enhanced time crystals and molecular knots}
\author{Jin Dai}
\email{daijing491@gmail.com}
\affiliation{Nordita, Stockholm University, Roslagstullsbacken 23, SE-106 91 Stockholm, Sweden}
\author{Xubiao Peng}
\email{xubiaopeng@bit.edu.cn}
\affiliation{Center for Quantum Technology Research and School of Physics, 
Beijing Institute of Technology, Beijing 100081, P. R. China}
%
\author{Antti J. Niemi}
\email{Antti.Niemi@su.se}
\affiliation{
Laboratoire de Mathematiques et Physique Theorique
CNRS UMR 6083, F\'ed\'eration Denis Poisson, Universit\'e de Tours,
Parc de Grandmont, F37200, Tours, France}
\affiliation{Nordita, Stockholm University, Roslagstullsbacken 23, SE-106 91 Stockholm, Sweden}
\affiliation{School of Physics, Beijing Institute of Technology, Haidian District, Beijing 100081, People's Republic of China}
\affiliation{Laboratory of Physics of Living Matter, Far Eastern Federal University, 690950, Sukhanova 8, Vladivostok, Russia}

\maketitle

\noindent
{\large
\bf 
A  time crystal is a time dependent physical  system  that 
does not reach a standstill, even in state of minimum energy.  
Here we show that the stability of a
time crystal can be enhanced by its topology. For this we simulate 
time crystals made of chainlike ensembles of mutually interacting point particles. 
When we tie the chain into a knot we find that its timecrystalline qualities improve.
The theoretical models we consider are widely used in coarse grained descriptions of linear 
polymers. Thus we expect that physical realizations of time crystals can be found in terms of knotted molecules.
}

\vskip 0.5cm

A classical Hamiltonian time crystal is a time dependent, time periodic solution of Hamilton's equation of motion that is simultaneously a  local minimum of its free
energy \cite{Wilczek-2012,Shapere-2012,Li-2012,Sacha-2018,Nayak-2019,Elze-2019}. Isolated, energy conserving
Hamiltonian time crystals were initially thought to be  impossible \cite{bruno-2013,Watabane-2015}. However,  explicit  timecrystalline solutions have been
recently constructed in effective theory models that describe a small number of pointlike interaction centers with dynamics derived from 
degenerate Poisson brackets \cite{Dai-2019}.  Here we employ such pointlike interaction centers to build linear objects akin a cyclic
molecular chain. We show that the  timecrystalline character of the chain improves once we tie it into a knot.

Knotted molecules and other topologically elaborate molecular structures have many remarkable  physical and chemical properties, from selective ion binding 
and strong catalytic activity to intricate molecular machines \cite{Dietrich-1989,Lukin-2005,Sauvage-2008,Erbas-2015,Marcos-2016,Horner-2016,Fielden-2017,Segawa-2019} 
They  provide unique opportunities to construct entirely new materials with exceptional strength and elasticity.  Sometimes a knotted molecule can even 
undergo autonomous swirling motion, in the way of a molecular motor \cite{Segawa-2019}; see also \cite{Gruziel-2018}. 
Our results propose that  knotted molecules are also excellent  candidates for  constructing autonomous 
energy conserving, topologically stable Hamiltonian time crystals.
 
We use a coarse grained approach and depict a knotted molecule of covalently connected atoms 
as a discrete, piecewise linear chain.  
The vertices  $\mathbf x_i $  ($i=1,...,N$)  coincide with  the locations of the pointlike  atoms,  and the
links $\mathbf n_i = \mathbf x_{i+1} - \mathbf x_i$ model the  covalent  bonds. We assume 
that the chain is closed and for this we set  $\mathbf x_{N+1} = \mathbf x_1$. 
Since translation symmetry remains unbroken, the  links $\mathbf n_i$ are the appropriate dynamical coordinates.
We assume that their lengths  $|\mathbf n_i |$ are  fixed, like covalent bond lengths should be in a coarse grained approach. 

We build our energy conserving Hamiltonian dynamics on the ``gyropic'' Lie-Poisson brackets \cite{Dai-2019}
\begin{equation}
\{ n_i^a , n_j^b \} = - \epsilon^{abc} \delta_{ij} n^c_i 
\la{n-bra}
\end{equation}
Since $ \{ \mathbf n_i , \mathbf n_k \cdot \mathbf n_k \} =0 $ 
for all pairs ($i,k$) 
the brackets (\ref{n-bra}) retain all the link  lengths intact, independently of the Hamiltonian details.  Thus
the brackets (\ref{n-bra}) are designed to generate any kind of a local chain motion,  except for stretching and shrinking,
and with no loss of generality we set $|\mathbf n_i |= 1$.  

We impose the chain closure as a constraint,
\begin{equation}
\mathbf G = \sum_{i=1}^N \mathbf n_i =0
\la{G}
\end{equation}
and since $\{ G^a , G^b \} = -\epsilon^{abc} G^c$
the constraint is of first-class. 
With $H(\mathbf n)$ our Hamiltonian free energy function, the Hamiltonian equation of motion that follows from the brackets (\ref{n-bra}) is
\begin{equation}
\frac{ \partial \mathbf n_i}{\partial t} = \{  \mathbf n_i , H  \} 
= - \mathbf n_i \times  \frac{\partial H}{\partial \mathbf n_i}  
\la{eom}
\end{equation}
and whenever $H(\mathbf n)$ has (weakly) vanishing Poisson brackets with (\ref{G}), 
an initially closed chain remains closed under the time evolution (\ref{eom}).

Time translation is a symmetry of (\ref{eom}). But if a solution can be found 
that is simultaneously both  time {\it dependent} and a minimum of $H(\mathbf n)$,  time translation symmetry becomes spontaneously broken. 
A {\it time crystal} is a time dependent and
periodic $\mathbf n_i(t+T) = \mathbf n_i(t)$ minimum energy solution of (\ref{eom}).

Here we do not elaborate on the physical conditions under which the equation (\ref{eom})  provides an effective theory description of molecular 
chain dynamics.  This is a question that should be resolved by comprehensive all-atom quantum molecular dynamics analyses of actual molecular chains
under proper ambient conditions.  It suffices to note that  Poisson brackets such as (\ref{n-bra}) 
are designed to describe self-induced motion in a general context \cite{Rasetti-1973,Holm-2003} and we refer to \cite{Segawa-2019,Gruziel-2018} 
for conceivable, recently proposed experimental scenarios, in the present case of molecular knots.

The Hamiltonians we consider  include  the  following two  free energy functions  \cite{Dai-2019} 
\begin{equation}
H_1=   a \sum\limits_{i=1}^N \mathbf n_i \! \cdot \! \mathbf n_{i+1} 
\ \ \ \ \ \& \ \ \ \ \
H_2 = b \sum\limits_{i=1}^N  \mathbf n_i \! \cdot \! (\mathbf n_{i+1} \! \times \! \mathbf n_{i-1}) 
\la{H12}
\end{equation}
Both have vanishing Poisson brackets with the constraints (\ref{G}). The Hamiltonian $H_1$ is akin the worm-like chain free energy of bending, it
is widely used in studies of  (bio)polymers and molecular chains \cite{Marko-1994,Marko-1995} and $H_2$ extends it to include a coupling between bending and twisting. 
To introduce additional Hamiltonian functions, we write the  vector that connects  
any two vertices $\mathbf x_i$ and $\mathbf x_j$ (with $i>j$) of our closed chain  
in the following symmetrized fashion in terms of  the links $\mathbf n_i$,
\begin{equation}
\mathbf x_i - \mathbf x_j  
=  \frac{1}{2} ( \mathbf n_j + ... + \mathbf n_{i-1}  - \mathbf n_{i} - ... - \mathbf n_{j-1}) 
\la{rn}
\end{equation}
Since the Poisson brackets between the distances $|\mathbf x_i - \mathbf x_j|$ and the constraint functionals
(\ref{G}) vanish, we may include in our Hamiltonian any two-body interaction  potential energy $V(|\mathbf x_i - \mathbf x_j|)$.   
An example is the Coulomb interaction
\begin{equation}
U(\mathbf x_1,...,\mathbf x_N)  = \frac{1}{2}  \underset{i\not= j}{\sum_{i,j=1}^{N}}  \frac{ e_i e_j  } { |\mathbf x_{i} - \mathbf x_j | }
\la{U}
\end{equation} 
where $e_i$ is the charge at the vertex $\mathbf x_i$.
Note that since the bond lengths are preserved by the  bracket (\ref{n-bra})  the nearest-neighbor contributions only add a constant  to the energy function. 

For  a more realistic description of a molecule, we may also include the attractive van der Waals  and the repulsive Pauli exclusion interactions between vertex pairs,
these are commonly described  by the Lennard-Jones potential. At long distances the van der Waals interaction is minuscule in comparison to the Coulomb interaction.
Since  our goal is to present a proof-of-concept for knotted molecular time crystals, we do not  aim for a detailed analysis of a particular molecular structure, 
for clarity of presentation we only consider the Pauli exclusion
\begin{equation}
V( \mathbf x_1,...,\mathbf x_N)  =  \frac{1}{2}  \underset{ i\not= j}{\sum_{i,j=1}^{N}}  \left(\frac{ r_{min} } { |\mathbf x_{i} - \mathbf x_j | }\right)^{\!\! 12}
\la{V}
\end{equation}
This introduces self-avoidance and prevents chain crossing;
in the case of actual molecules two covalent bonds are not supposed to cross each other.  

We note that Hamiltonian functions that  are combinations of  (\ref{H12}), (\ref{U}) and (\ref{V}) commonly appear as a free energy in coarse grained, effective theory descriptions 
of linear molecules and polymer chains,
at relatively large distance and time  scales where all local details of atomic structure can be overlooked \cite{Marko-1994,Marko-1995}.

We start our search  for a time crystal  by locating a minimum 
energy configuration of the Hamiltonian. For this we amend the equation (\ref{eom})
with  a diffusion term,
\begin{equation}
\frac{ \partial \mathbf n_i}{\partial t}  
= - \mathbf n_i \times  \frac{\partial H}{\partial \mathbf n_i}  +  \mu \, \mathbf n_i \times (\mathbf n_i \times  
\frac{\partial H}{\partial \mathbf n_i} )
\la{eom2}
\end{equation}
where $\mu \geq 0$ is the diffusion coefficient. When $\mu \not=0$ (\ref{eom2}) no longer 
preserves the constraint (\ref{G}) and we need to enforce it explicitly. For this we deploy a 
Lagrange multiplier ${\boldsymbol \lambda} $ that invokes the constraint (\ref{G}) as an
equation of motion. Accordingly,  we extend the Hamiltonian into $H \to H_{\boldsymbol \lambda} = H + {\boldsymbol \lambda} \cdot \mathbf G$
and when  we substitute this in  (\ref{eom2}) we obtain the following equation,
\begin{equation}
\frac{d H_{\boldsymbol \lambda}}{dt} = - \frac{\mu}{1+\mu^2}\sum\limits_{i=1}^N \left |  \frac{ d \mathbf n_i }{d t} \right |^2
\la{gilbert}
\end{equation}
The 
time evolution (\ref{eom2})
proceeds towards decreasing values of  $ H_{\boldsymbol \lambda}$ and the flow
continues until a stable local minimum ($\mathbf n_0, {\boldsymbol \lambda}_0$) of $H_{\boldsymbol \lambda}$ has been reached.
The Lagrange multiplier theorem \cite{Abraham-1988} states that $\mathbf n_0$ minimizes $H(\mathbf n)$
on the constraint surface (\ref{G}) and we can resolve for $ {\boldsymbol \lambda}_0$  as a function of $\mathbf n_0$,  with result $
{\boldsymbol \lambda}_0 = - \frac{ \partial H} {\partial \mathbf n_i }_{| \mathbf n_0}$

Thus, on  the constraint surface (\ref{G}) we can  express the equation (\ref{eom}) as 
\[
\frac{ \partial \mathbf n_i}{\partial t}  = {\boldsymbol \lambda}_0 \times  \mathbf n_i 
\]
together with $\sum_{i=1}^N \mathbf n_i =0$.
Whenever ${\boldsymbol \lambda}_0  \not=0$ we then
have a time crystal that rotates like a rigid body, with the direction 
of rotation and the angular velocity  both 
determined by  (time independent) ${\boldsymbol \lambda}_0 $.

\vskip 0.2cm
We argue that time crystals  are common in knotted molecules. 
For this it suffices to describe examples that have the topology of a trefoil knot, as knotted time crystals  with a more complex topology
can be constructed  similarly. 
To construct an initial {\it Ansatz} trefoil for the flow equation (\ref{eom2}), we start with the continuum  trefoil 
 \begin{equation}
~~~~  \begin{matrix}\vspace{0.1cm}
x_1(s) &=& L \cdot [\, \cos(s) - A \cos(2s)] \\ \vspace{0.1cm}
x_2(s) &=& L \cdot [\,  \sin(s) + A \sin(2s)] \\
x_3(s) &=&  \pm \, L \cdot [\,  \sqrt{1+A^2}\sin(3s) ]
 \end{matrix}  \ \ \ \ \ \ s \in [0,2\pi)
  \la{3foil}
 \end{equation}
Here $L$ and $A$ are parameters 
and $\pm$ determines whether the trefoil is left-handed (+) or right-handed (-).  This trefoil has a high level of three-fold symmetry, for example each of the
three coordinates has the radius of gyration value $R_g=  L \sqrt{1+A^2} $.  

Due to the long distance interactions (\ref{U}) and (\ref{V}),  the computer time that is needed to reach the energy minimum as a solution of 
the flow equation (\ref{eom2}) grows rapidly with the number of
vertices. For that reason, in all our examples we have only $N=12$ vertices.  
To discretize (\ref{3foil}) accordingly, we first divide it 
into three segments,  all with an equal parameter length $\Delta s = 2\pi/3$.  We then divide each of these three
segments  into four subsegments, all with an equal length in space for $N=12$ vertices. 
We set  $A=2$ and when we choose $L=0.340$ each segment has a unit length, and the
three space coordinates ($x_1,x_2,x_3$) have the radius of gyration 
\begin{equation}
R_g^{(i)} \ = \ \sqrt{ \frac{1}{N} \sum_{n=1}^N ( x_i(n)- \bar x_i)^2}
\la{Rg}
\end{equation} 
values ($0.722, 0.722, 0.715$); here $\bar x_i$ is the average of the $x_i(n)$.
This constitutes the initial  discrete trefoil  {\it Ansatz} that we use in the flow equation (\ref{eom2}), and we now describe three representative time crystal solutions:

As a first example, we consider a Hamiltonian that is  a linear combination of the  bending rigidity $H_1$ in (\ref{H12}) and the  Pauli exclusion (\ref{V}), with  the parameter values
$a=1/4$ and $r_{min} =  3/4$. The flow (\ref{eom2})  terminates to a minimal energy trefoil
with radius of gyration values ($0.710, 0.710, 0.874$); note that the initial {\it Ansatz} trefoil is slightly oblate in the $x_3$ direction, but the minimal energy trefoil is slightly prolate. 
We set $\mu=0$  to confirm that we have a time crystal that rotates around the $x_3$ axis
with angular velocity $\omega \approx 0.619$ in our units;   the  direction of rotation depends on 
the sign of  $x_3$ in (\ref{3foil}).   

Our second  example is  a  sum of the twist-bend coupling $H_2$ in (\ref{H12}) and the Pauli exclusion
(\ref{V}), with parameter values $b=1/4$ and $r_{min} =  3/4$. 
Now the  flow (\ref{eom2}) terminates at a minimum energy prolate trefoil,
with radius of gyration values ($0.715, 0.715, 0.875$). When we set $\mu=0$  we find that this trefoil
is a time crystal that rotates around the $x_3$ axis with angular velocity $\omega \approx 1.046$. 

In the third example  the Hamiltonian is  a sum  of the Coulomb interaction (\ref{U}) and  the Pauli exclusion (\ref{V}), with parameters $e_i=1$ and  $r_{min} =  3/4$.
The flow (\ref{eom2}) terminates at a prolate trefoil with radius of gyration values ($0.717, 0.717, 0.889$). With $\mu=0$ we again have
a time crystal, now with angular velocity $\omega \approx 1.571$.  

Note that we do not account for energy loss due to electromagnetic radiation effects, in the case of a Coulomb interaction. We assume that radiation effects are
minor, at a time scale that is pertinent to our analysis.

\begin{figure}[h]
        \centering
                \includegraphics[width=0.85\textwidth]{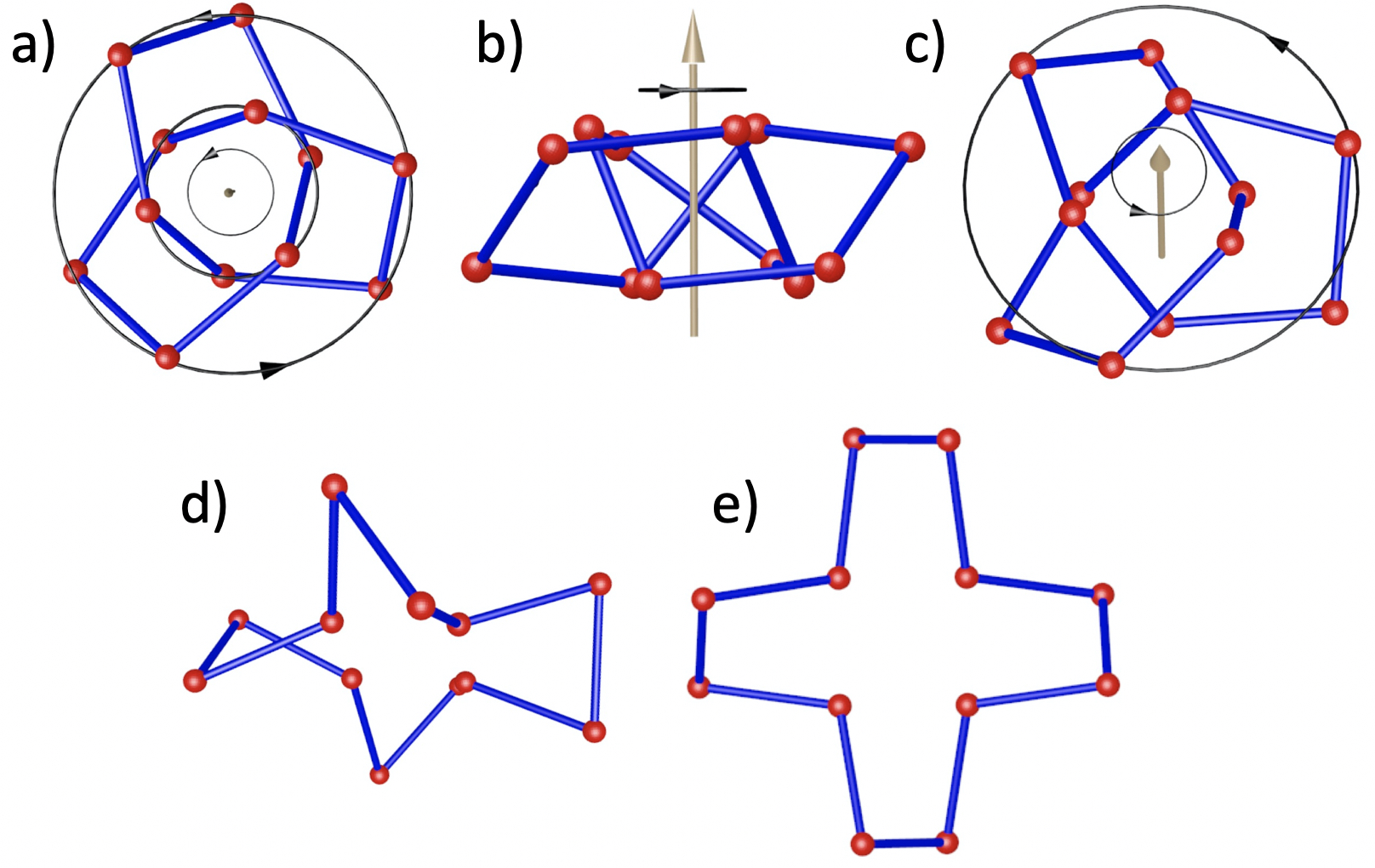}
        \caption{ \small Panels a)-c) show the minimum energy time crystal sol\-u\-tion of (\ref{eom}) with Hamiltonian that is a combination of (\ref{U}) and  (\ref{V}).
      Panel a) is a view along the rotation  axis $x_3$. It shows the  presence of a 3-fold 
      spatial crystalline symmetry. Panel b)  is a viewpoint that is normal to the  rotation axis,  revealing additional spatial crystalline symmetry.  Panel c) shows a generic
      view.  In a) and c)  the black lines with arrows depict how the time crystal rotates. Panels d) and e):  The  minimum energy unknotted 
      solution of (\ref{eom}) with (\ref{U}) and  (\ref{V}).  Panel d)  shows a generic view while e) is a view from the top, displaying the 4-fold symmetry. The structure rotates {\it very} slowly around the
      4-fold symmetry axis.}
       \label{fig-1}
\end{figure}

Notably, all the three timerystalline trefoils have a very similar slightly prolate shape, all the mutual root-mean-square distances between them
are less that 0.01. In Figure \ref{fig-1} panels  a)-c) we illustrate  the  third  as an example and depict the way it rotates. 
%
%
%
%
%
%
%
%
%
%
%
%
%
%
%
%
%
%
%
%
%
%
%
%
%
%
We observe that its structure displays a remarkable three-fold spatial crystalline symmetry. 

We argue that the  knotted  time crystals we have constructed,  exemplify a  general pattern:  When the knottiness of a chain  increases, so does  its timecrystalline 
character.  The reason is that our Hamiltonians  are highly symmetric and thus their critical point sets are also symmetric,  and the topological constrains a knot
imposes on the shape of the structure  are often inconsistent with these symmetries. This mismatch between the critical point set of
the Hamiltonian and the topology of the knot causes a frustration that drives the  timecrystalline dynamics.

For example, in the case of an unknot,  for $a>0$ the Hamiltonian $H_1$  acquires an absolute minimum at the critical point where all $\mathbf n_i \cdot \mathbf n_{i+1} = -1$ 
and for these values the solution of (\ref{eom}) is stationary, there is no time crystal.
Similarly, when  $a$ is negative the regular, fully symmetric planar dodecagon is the unknotted energy minimum of the Hamiltonian $H_1$, and this structure displays no time crystalline 
dynamics when we substitute it into (\ref{eom}). 
The  regular dodecagon is also  the unknotted 
minimum energy configuration of the Coulomb potential (\ref{V}) when the $e_i$  are positive, and again there is no time crystal solution of (\ref{eom})  
Thus, in all our  examples the existence and stability of the time crystal solution that we have constructed is entirely due to the trefoil knot topology.

Finally, in the case of the Hamiltonian $H_2$ the timecrystalline dynamics is intriguing even in the case of an unknotted chain: It has been shown \cite{Dai-2019} that  for $N=3$ the 
Hamiltonian $H_2$ gives rise to a time 
crystal in the shape of an equilateral triangle, and for $N=4$ it supports a time crystal that relates to the tetragonal disphenoid. 
In the present case, with  $N=12$,  the flow (\ref{eom2}) terminates  
in the jagged unknot shown in Figure  \ref{fig-1} panels d) and e).
In this unknotted minimum energy configuration the vectors $\mathbf n_i$ and $\partial H / \partial \mathbf n_i$ acquire their maximally antiparallel orientation 
that is allowed by the chain closure constraint; we compute 
\begin{equation}
 \arccos \left( \frac{ \mathbf n_i \cdot  \frac{\partial H}{ \partial \mathbf n_i}  }{ ||  \frac{\partial H}{ \partial \mathbf n_i} ||  }\right) 
 \ \approx  \ 3.119 \ \  {\rm (rad)}
\label{angle}
\end{equation}
When we set $\mu=0$ the structure shown in Figure  \ref{fig-1} panels d), e)  is a time crystal that rotates around the four-fold 
symmetry axis of  panel e)  but with a {\it very} small angular velocity $\omega \approx 0.0158$.  More generally, 
when the number of vertices $N$ increases the angles (\ref{angle}) 
approach the value $\pi$, where the time crystal comes to a stop.  But if  the chain then forms a knot its timecrystalline dynamics recovers.
Thus we can control the angular velocity of a time crystal, both by adjusting the number of vertices (atoms) and by adjusting
the level of knottiness. This ability to  control the dynamics  of a time crystal could
be most valuable {\it e.g.} in the construction of timecrystalline precision clocks.

\vskip 0.2cm
In summary, topology has a pivotal role in determining  the timecrystalline character of a closed molecular chain.
In particular,  we have found that even when an unknotted chain can not  support any  time crystal, it often becomes timecrystalline when we tie it into a knot.
The examples that we have analyzed have the topology of a trefoil knot, but we are confident that our conclusions extend 
to more general knotted topologies.   Moreover, the Hamiltonians that we have studied are quite universal. They are employed  widely in coarse grained effective theory 
descriptions of (bio)polymers and chain molecules. Thus we expect that  actual physical  realizations of our  topological, knotted time crystals can be
found in terms of actual material systems. For the identification of promising molecular candidates, at the level of  actual chemical composition, 
one needs to perform simulations with more realistic energy  functions.  For example, one can use the force fields  that are employed in all atom molecular dynamics.
However,  the computational challenges posed by such simulations are arduous, without a cut-off for the range of long distance interactions the energy minimization takes a
very long  time with presently available computers. Thus we defer these investigations  to the  future.  But once found and experimentally 
constructed, topologically stable molecular time crystals  should find various applications, from high precision time measurement to molecular motors.

\vskip 0.3cm

\noindent
Acknowledgements: JD and AJN thank Frank Wilczek  for discussions. The work by JD and AJN has been supported by the Carl Trygger Foundation, by the 
Swedish Research Council under Contract No. 2018-04411, and by COST Action CA17139. The work by XP is supported by Beijing Institute of Technology 
Research Fund Program for Young Scholars.

\end{document}